\documentstyle[11pt]{article}

\begin{document}

\title{{\bf Non-integrability and Mach's principle in Induced Matter Theory}}
\author{S. Jalalzadeh$^1$\thanks{
email: s-jalalzadeh@sbu.ac.ir} \\
$^1${\small Department of Physics, Shahid Beheshti University, Evin, Tehran
19839, Iran}\\
}
\maketitle

\begin{abstract}
The geodesic equation in induced matter theory is redefined. It is
shown that the extra forces do not change the four-impulse of
massive particles. We show that the $4D$ space-time is
non-integrable and  find the relation between non-integrability
and the Mach's principal. \vspace{5mm}\newline PACS numbers:
04.50.+h, 04.20.Cv
\end{abstract}

\vspace{2cm}

\section{Introduction}

Matter or the source of spacetime and fields are basic concepts of classical
field theories, so that the Einstein tensor is expressed in terms of the
geometry of spacetime and matter is expressed by the corresponding
energy-momentum density tensor. This two basics are connected by the
Einstein field equation
\begin{equation}
G_{\mu\nu} = -8\pi G_N T_{\mu\nu}.  \label{0}
\end{equation}
According to equation (\ref{0}) distribution of matter determines
the geometric properties of spacetime. On the other hand, one can
read Einstein's equation in the opposite direction, and expect
creation of matter by geometry. One of the Einstein's dreams was
to build a gravitational theory in which the concept of matter is
rejected in favour of pure fields \cite{Einstein0}. A
gravitational theory in which the matter is absorbed into the
field itself, so that one has a set of homogeneous partial
differential equations according to Einstein \cite{Einstein} is
called unified field theory. In this kind of theories it is well
known that the Mach's principle is satisfied \cite{Callaway}.
There exist various extensions of Einstein's framework extracting
matter from pure geometry. One of the interesting extensions of
Einstein's theory is based on the idea that our $4D$ spacetime
(brane) is a submanifold embedded in a higher dimensional manifold
(bulk). The origin of this idea is in Kaluza and Klein papers that
proposed a unification of electromagnetism and gravity. In
Kaluza-Klein (KK) theory, the extra dimension plays a purely
formal role and the components of the $5D$ metric tensor do not
depend on the extra coordinate. The idea of extra dimensions in
which ordinary matter is confined within a brane, has attracted an
enormous amount of attention during the last few years. The early
works in this approach was done by Maia \cite{Maia1}, Joseph
\cite{Joseph},
Akama \cite{Akama}, Rubakov,  Shaposhnikov \cite{Rubakov} and Visser \cite%
{Visser}. A revised Kaluza-Klein approach in the direction of
unified field theory is based on Wesson's theory \cite{Wesson1},
in which matter on brane is created by the geometry of the bulk
space. This theory is different from the classical KK scenario by
the fact that it has a noncompact fifth dimension and that the
$5D$ bulk space is devoid of matter. For this reason it is called
induced matter theory (IMT) where the effective $4D$ matter is a
consequence of the geometry of the bulk. That is, in IMT, the $5D$
bulk space is Ricci-flat while the $4D$ hypersurface is curved by
the $4D$ induced matter. In this paper we will show how Mach's
principle can be interpreted according to IMT theory. Another
extension of general relativity  that can generate massive
particles is the Weyl-Dirac theory \cite{Dirac}. In this approach
the Weyl vector field can be regarded  as the creator of matter
fields \cite{Isra}. The Weyl-Dirac theory is based on the
non-integrable geometry or Weylian geometry. In Weylian geometry
the component of a vector and its length changes under parallel
displacement. The variation of length is a consequence of the Weyl
vector field and is similar to the generation of massive matter
fields. We will show that the length of $4$-vectors in IMT as in
the Weylian geometry changes under a parallel displacement. The
variation of mass and length  are both consequences of the motion
of test particle's in the direction of extra dimension, so that
from the point of view of a $4D$ observer these variations are
consequences of Mach's principle.

The organization of paper is as follows: In section 2 we describe the
dynamics of test particles. Section 3 deals with the non-integrability of
induced $4$-vectors and in Section 4 we present induced mass and mach's
principle. Conclusions are drawn in the last section.

\section{Test particle dynamics}

In IMT theory, we assume that our spacetime can be isometrically
and locally embedded in a Ricci-flat $5D$ spacetime. In contrast
to the Randall and Sundrum brane models where the matter field is
confined to the fixed brane, in IMT there is no mechanism in IMT
to confine induced matter field exactly on a specific brane. The
authors of \cite{Seahra} and \cite{Jalal} show  that to confine
test particles on a brane it is necessary to either a
non-gravitational centripetal confining force, with an unknown
source, or assume that our brane is totaly geodesic in which case
it is impossible to embed an arbitrary brane in the bulk space. In
IMT however, if the induced matter field
satisfies  ``machian strong energy condition'' $T_{\mu \nu }u^{\mu }u^{\nu }+\frac{1%
}{2}T>0$, and also we have more than one extra dimension, then the
test particles becomes stable around the fixed brane
\cite{Jalal2}. This means that, the motion of test particles along
the extra dimensions becomes very limited (for ten extra
dimensions, it is of order  $10^{-12}cm$) and we can not see
disappearance of particles at ordinary scales of energy. But, in
the literature it is common to assume that one extra dimension is
enough to obtain any kind of induced matter \cite{Wesson1}. To
solve this problem according to \cite{Goshtas} one can assume that
the speed of propagation in the $5D$ bulk space is  much greater
than on the brane and consequently the size of the fluctuations of
test particle in the bulk space becomes very limited and we have
stabilized test particles. Finally, we can say that in  IMT at
the large scales we have matter field confined to  a fixed brane, say $%
\bar{g}_{\mu \nu }$ that satisfies induced Einstein field equations \cite%
{Jalal2}, and at small scales we find the matter fields having
small fluctuations around this brane. If we denote the metric of this brane by $%
g_{\mu \nu }$, then it becomes acceptable to assume that this new
brane is a perturbation of the original one $\bar{g}_{\mu \nu }$
\cite{Jalal2}. In the following we briefly review the relation of
geometrical objects in these two branes.

Consider the background manifold $\overline{V}_{4}$ isometrically embedded
in ${V}_{5}$ by a map ${\cal Y}:\overline{V}_{4}\rightarrow V_{5}$ such that
\begin{equation}
{\cal G}_{AB}{\cal Y}_{\,\,\,,\mu }^{A}{\cal Y}_{\,\,\,,\nu }^{B}=\bar{g}%
_{\mu \nu },\hspace{0.5cm}{\cal G}_{AB}{\cal Y}_{\,\,\ ,\mu }^{A}{\cal N}%
^{B}=0,\hspace{0.5cm}{\cal G}_{AB}{\cal N}^{A}{\cal N}^{B}=1  \label{mama41}
\end{equation}%
where ${\cal G}_{AB}$ $(\bar{g}_{\mu \nu })$ is the metric of the bulk
(brane) space $V_{5}(\overline{V}_{4})$ in an arbitrary coordinate with
signature $(-,+,+,+,+)$, $\{{\cal Y}^{A}\}$ $(\{x^{\mu }\})$ are the basis
of the bulk (brane) and ${\cal N}^{A}$ is a normal unit vector orthogonal to
the brane. Perturbation of $\overline{V}_{4}$ in a sufficiently small
neighborhood of the brane along an arbitrary transverse direction $\zeta$ is
given by
\begin{equation}
{\cal Z}^{A}(x^{\mu },\xi)={\cal Y}^{A}+({\cal L}_{\zeta }{\cal Y})^{A},
\label{B}
\end{equation}%
where ${\cal L}$ represents the Lie derivative and $\xi$ is a small
parameter along ${\cal N}^{A}$ parameterizing the extra noncompact
dimension. By choosing $\zeta$ orthogonal to the brane we ensure gauge
independency \cite{Jalal} and have perturbations of the embedding along a
single orthogonal extra direction $\bar{{\cal N}}$, giving the local
coordinates of the perturbed brane as
\begin{equation}
{\cal Z}_{,\mu }^{A}(x^{\nu },\xi)={\cal Y}_{,\mu }^{A}+\xi\bar{{\cal N}}%
_{\,\,\,,\mu }^{A}(x^{\nu }).  \label{C}
\end{equation}%
In a similar manner, one can find that since the vectors $\bar{{\cal N}}%
^{A}$ depend only on the local coordinates $x^{\mu }$, they do not
propagate along the extra dimension. The above assumptions lead to
the embedding equations of the perturbed geometry
\begin{equation}
{\cal G}_{\mu \nu }={\cal G}_{AB}{\cal Z}_{\,\,\ ,\mu }^{A}{\cal Z}_{\,\,\
,\nu }^{B},\hspace{0.5cm}{\cal G}_{\mu 4}={\cal G}_{AB}{\cal Z}_{\,\,\ ,\mu
}^{A}{\cal N}^{B},\hspace{0.5cm}{\cal G}_{AB}{\cal N}^{A}{\cal N}^{B}={\cal G%
}_{44}.  \label{E}
\end{equation}%
If we set ${\cal N}^{A}=\delta _{4}^{A}$, the line element of the bulk space
in the Gaussian frame (\ref{E}) is given by
\begin{equation}
dS^{2}={\cal G}_{AB}d{\cal Z}^{A}d{\cal Z}^{B}=g_{\mu \nu
}(x^{\alpha },\xi )dx^{\mu }dx^{\nu }+d\xi ^{2},  \label{2}
\end{equation}
where
\begin{equation}
g_{\mu \nu }=\bar{g}_{\mu \nu }-2 \xi\bar{K}_{\mu \nu }+\xi^2 \bar{g}%
^{\alpha \beta }\bar{K}_{\mu \alpha}\bar{K}_{\nu \beta},  \label{G}
\end{equation}%
is the metric of the perturbed brane, so that
\begin{equation}
\bar{K}_{\mu \nu}=-{\cal G}_{AB}{\cal Y}_{\,\,\,,\mu }^{A}{\cal N}_{\,\,\
;\nu }^{B},  \label{H}
\end{equation}
represents the extrinsic curvature of the original brane (second
fundamental form). Any fixed $\xi$ signifies a new perturbed
brane, enabling us to define an extrinsic curvature similar to the
original one by
\begin{equation}
{K}_{\mu \nu}=-{\cal G}_{AB}{\cal Z}_{\,\,\ ,\mu }^{A}{\cal N}_{\,\,\ ;\nu
}^{B}=\bar{K}_{\mu \nu }-\xi\bar{K}_{\mu \gamma }\bar{K}_{\,\,\ \nu
}^{\gamma }.  \label{J}
\end{equation}

Following this section we wish to derive the $4D$ geodesic
equation from a $5D$ Lagrangian to obtain the acceleration
equation induced in $4D$. To study the dynamics in $4D$ IMT
gravity, we begin by extremizing the following action which is
equivalent to the usual action for a test particle
\begin{equation}
{\cal I}=\frac{1}{2}\int_{\lambda _{A}}^{\lambda ^{B}}d\lambda \left[
e^{-1}(\lambda ){\cal G}_{AB}\dot{{\cal Z}}^{A}\dot{{\cal Z}}%
^{B}-M^{2}e(\lambda )\right] ,  \label{1}
\end{equation}%
where $\lambda $ is an arbitrary parameter on the worldline with endpoints $%
A $ and $B$, $e(\lambda )$ the \textquotedblleft
einbein\textquotedblright\ is a new function and $M$ is the
particle mass in the bulk space. According to  \cite{Romero} there
is two main approaches in obtaining induced geodesic equation on
the brane. This approaches are the so-called foliating and
embedding methods. In this paper we use the first (foliating)
approach. The Variation of the action (\ref{1}) with respect to
$e$ and ${\cal Z}^{A}$ lead to
\begin{equation}
e=\frac{1}{M}\sqrt{{\cal Z}_{AB}\frac{d{\cal Z}^{B}}{d\lambda }\frac{d{\cal Z%
}^{A}}{d\lambda }}=\frac{1}{M}\frac{dS}{d\lambda },  \label{3}
\end{equation}%
and
\begin{equation}
\frac{d\dot{{\cal Z}}^{A}}{d\lambda }+\bar{\Gamma}_{BC}^{A}\dot{{\cal Z}}^{B}%
\dot{{\cal Z}}^{C}=\frac{\dot{e}}{e}\dot{{\cal Z}}^{A},  \label{4}
\end{equation}%
where an overdot represents derivative with respect to $\lambda $ and $\bar{%
\Gamma}_{BC}^{A}$ denotes the Christoffel symbols of the bulk
space. To understand the nature of the induced motion we must
project the geodesic equation (\ref{4}) onto the brane. The
resulting equations that describes equations of motion on the
brane and orthogonal to the brane respectively become
\begin{equation}
\dot{u}^{\mu }+\Gamma _{\alpha \beta }^{\mu }u^{\alpha }u^{\beta }=\frac{%
\dot{e}}{e}u^{\mu }+2K_{\,\,\,\alpha }^{\mu }u^{\alpha }u^{4},  \label{5}
\end{equation}%
and
\begin{equation}
\dot{u^{4}}+K_{\alpha \beta }u^{\alpha }u^{\beta }=\frac{\dot{e}}{e}u^{4},
\label{6}
\end{equation}%
where $\Gamma _{\alpha \beta }^{\mu }$ is the the Christoffel symbols of the
brane, $u^{\mu }=dx^{\mu }/d\lambda $, $u^{4}=d\xi /d\lambda $ and $K_{\mu
\nu }=-\frac{1}{2}g_{\mu \nu ,\xi }$ denote the extrinsic curvature. In the
above equations of motion the \textquotedblleft einbein\textquotedblright\
function is unknown and is related to parametrization of the path of test
particles. Using the explicit form of the line element of the bulk space (\ref{2}%
) and equation (\ref{3}) we have
\begin{equation}
{l}^{2}+(u^{4})^{2}=e^{2}M^{2},  \label{7}
\end{equation}%
where ${l}^{2}:=g_{\mu \nu }u^{\mu }u^{\nu }$. At this stage the
parametrization of the worldline is arbitrary and for this reason
the norm of the $4$-velocity of the particle is not normalized to
unity. Now, differentiating equation (\ref{7}) and using equation
(\ref{6}) we obtain
\begin{equation}
\frac{\dot{l}}{l}=\frac{\dot{e}}{e}+\frac{1}{l^{2}}K_{\alpha \beta
}u^{\alpha }u^{\beta }u^{4}.  \label{8}
\end{equation}%
The usual assumption in the literature {\cite{Wesson2}} on the
parametrization of the path of test particles is
\begin{equation}
d\lambda ^{2}=-ds^{2}:=-g_{\mu \nu }dx^{\mu }dx^{\nu }.  \label{9a}
\end{equation}%
We then have $u^{\mu }u_{\mu }=-1$, or in other words $l^{2}=-1$.
In this case the variation of the magnitude of the $4$-velocity
vanishes and the result from equation (\ref{8}) becomes
\begin{equation}
\frac{\dot{e}}{e}=-K_{\alpha \beta }u^{\alpha }u^{\beta }u^{4},  \label{9}
\end{equation}%
Note that in this equation an overdot represents derivative
respect to the $s$ and $u^{\mu }=\frac{dx^{\mu }}{ds}$. Inserting
the above relation into equation (\ref{5}) the result is
\begin{equation}
\dot{u}^{\mu }+\Gamma _{\alpha \beta }^{\mu }u^{\alpha }u^{\beta }=f^{\mu },
\label{10a}
\end{equation}%
where $f^{\mu }$ is the acceleration of unit mass and is given by
\begin{equation}
f^{\mu }:=(2g^{\mu \nu }-u^{\mu }u^{\nu })K_{\alpha \nu }u^{\alpha }u^{4}.
\label{10b}
\end{equation}%
The last two equations show that the orthogonal part of
acceleration $f_{\mu }u^{\mu }$ does not vanish at all. On the
other hand it is a well known fact that all the $4D$ basic forces
lead to acceleration that are orthogonal to the $4$-velocity of
the particle \cite{Dicke}. Some authors \cite{Wesson2} related
this timelike acceleration to $4D$ physics by assuming that the
\textquotedblleft invariant" inertial mass of a test particle
varies along its worldline and the timelike acceleration
corresponds to this variation of inertial mass. Also in
\cite{Jalal2} this idea was generalized and shown that  variation
of inertial mass and charge can be related to the this normal
component of acceleration. Indeed the author of \cite{Ponce} tried
to show that this result is the same in both brane models and IMT,
i.e,., from observational viewpoint, the experiments measuring the
extra force acting on test particles are not able to discriminate
whether our universe is described by the brane world scenario or
by IMT. At this point, one can consider another possibility which
is different from the above approaches. Let us assume that the
magnitude of $4$-velocity of test particle is not an invariant  of
motion and the orthogonal part of the acceleration vanishes
$f^{\mu }u_{\mu }=0$. In this case, by contracting equations
(\ref{5}) with the $4$-velocity of a test particle and comparing
the result with (\ref{8}), we obtain
\begin{equation}
f^{\mu }=2(g^{\mu \nu }-\frac{1}{l^{2}}u^{\mu }u^{\nu })K_{\nu \alpha
}u^{\alpha }u^{4}.  \label{11}
\end{equation}%
\begin{equation}
\frac{\dot{e}}{e}=-\frac{2}{l^{2}}K_{\mu \nu }u^{\mu }u^{\nu }u^{4}.
\label{11b}
\end{equation}%
Also the variation of length of the $4$-velocity by inserting
equation (\ref{11b}) into the equation (\ref{8}) is given by
\begin{equation}
\frac{dl}{l}={\cal A}_{\beta }dx^{\beta },  \label{12}
\end{equation}%
where
\begin{equation}
{\cal A}_{\alpha }=-\frac{1}{l^{2}}K_{\alpha \beta }u^{\beta
}u^{4}=-K_{\alpha \beta }\frac{dx^{\alpha }}{ds}\frac{d\xi }{ds}.  \label{13}
\end{equation}%
In this case the above equations show that the normal component of
acceleration vanishes but on the other hand the magnitude of the
$4$-velocity of the test particle varies and consequently the
worldline of the test particle is not integrable. In the
Riemannian geometry the antisymmetry of the Riemann tensor in its
first pair of indices gives $\frac{1}{2}d(l^{2})=u_{\mu }du^{\mu
}=0$, i.e., the norm of the 4-velocity is integrable. One the
other hand, one obtains from equation (\ref{12}) the change in the
length of the 4-velocity that was transported round a closed loop
\begin{equation}
\triangle l\propto exp(\int F_{\mu \nu }dS^{\mu \nu }),  \label{13a}
\end{equation}%
where $dS^{\mu \nu }$ is the area element  and $F_{\mu \nu }={\cal
A}_{\mu ,\nu }-{\cal A}_{\nu ,\mu}$ is known as the length
curvature \cite{Perlic}. One concludes that the length is
non-integrable, unless ${\cal A}_{\mu }$ is a gradient vector, so
that $F_{\mu \nu }=0$ \cite{Isra1}. The interesting question at
this point is the meaning of the extra acceleration (\ref{11}). It
is well-known that in the non-integrable geometry the analogous
expression for the variation of mass is given by \cite{Rosen}
\begin{equation}
f^{\mu }=(g^{\mu \nu }-\hat{u}^{\mu }\hat{u}^{\nu })\frac{m_{,\nu
}}{m}, \hspace{1cm}\hat{u}^\mu = \frac{u^\mu}{l}.
\end{equation}%
The result is ${\cal A}_{\mu }$ is proportional to the variation
of inertial mass in this version of IMT. We will back to this
point in section 4. The summery of the above two approach is that
we can assume the worldline of particle is integrable and the
extra force has normal component, or the worldline is
non-integrable and consequently the extra force is similar to the
$4D$ basic forces and dose not have any normal component. In the
next section we will show that the second assumption seems to be
correct.

\section{Induced parallel displacement in IMT}

Consider an arbitrary vector in the $5D$ bulk space $X_A$ which
has a 4-dimensional counterpart on the brane the vector $X_\mu$.
This two vectors are related by
\begin{equation}
X_\mu = {\cal G}_{AB}X^A {\cal Z}^B_{,\mu}.  \label{14}
\end{equation}
Let us consider an infinitesimal parallel displacement of a vector in the
bulk space
\begin{equation}
dX_A = -\bar{\Gamma}^B_{AC}X_Bd{\cal Z}^C.  \label{15}
\end{equation}
The change of the length of this vector obviously vanishes. Now according to
equation (\ref{14}), the induced parallel displacement of $X_\mu$ is
\begin{equation}
dX_\mu = {\cal G}_{AM}\bar{\Gamma}^M_{BC}{\cal Z}^B_{,\mu}X^Ad{\cal Z}^C+
{\cal G}_{AB}X^Ad{\cal Z}^B_{,\mu}.  \label{16}
\end{equation}
As the bulk space may be mapped either by $\{{\cal Z}^A\}$ or by local
coordinates of brane and extra dimension $\{x^\mu,\xi\}$ one can write
\begin{equation}
d{\cal Z}^C = {\cal Z}^C_{,\alpha} dx^\alpha + {\cal N}^C d\xi.  \label{17}
\end{equation}
Inserting decomposition (\ref{17}) into the expression for the parallel
displacement (\ref{16}) we obtain
\begin{equation}
dX_\mu = {\cal G}_{AM}\left(\bar{\Gamma}^M_{BC}{\cal Z}^B_{,\mu}X^A +X^A
{\cal Z}^M_{,\mu,C} \right)\{ {\cal Z}^C_{,\alpha}dx^{\alpha} + {\cal N}^C
d\xi \}.  \label{18}
\end{equation}
In The Gaussian frame this may be rewritten as
\begin{equation}
dX_\mu = \Gamma^\beta_{\mu\alpha}X_\beta dx^\alpha + K_{\mu \alpha}X_4
dx^\alpha - K^\beta _{\,\,\mu}X_\beta d\xi,  \label{19}
\end{equation}
Now let us consider the square of the length $X^2 := g^{\mu\nu}X_\mu X_\nu$.
Its change under parallel displacement is
\begin{equation}
dX^2 = g^{\mu\nu}_{\,\,\,,\gamma}X_\mu X_\nu dx^\gamma +
g^{\mu\nu}_{,\xi}X_\mu X_\nu d\xi +2 g^{\mu\nu}X_\mu dX_\nu.  \label{20}
\end{equation}
Making use of $g^{\mu\nu}_{\,\,\,,\gamma} =
-\Gamma^\mu_{\gamma\beta}g^{\nu\beta} - \Gamma^\nu_{\gamma\beta}g^{\mu\beta}$
and $g^{\mu\nu}_{\,\,\,,\xi} = 2K^{\mu\nu}$, we obtain from equations (\ref%
{19}) and (\ref{20}) the change of the squared length of the $4$-vector
\begin{equation}
dX^2 = 2X^\mu X^4 K_{\mu\alpha}dx^\alpha.  \label{21}
\end{equation}
This equation is independent of the choice of model and is correct
both in brane models where  we have a fixed brane or in IMT where
matter lives  in the  perturbed brane $g_{\mu\nu}$. Thus, in
general case, the brane possesses a non-integrable geometry, and
only when the original $5D$ vectors do not have extra components,
or when the extrinsic curvature vanishes one has a Riemannian
brane. In the particular case if we set
\begin{equation}
X^A = \frac{d{\cal Z}^A}{dS} = {\cal Z}^A_{\,\,,\alpha}\frac{dx^\alpha}{dS}
+ {\cal N}^A\frac{d\xi}{dS},  \label{22}
\end{equation}
where $S$ is an affine parameter in the bulk space, then according
to equations (\ref{3}) and (\ref{14}) we have
\begin{equation}
X_\mu = g_{\alpha\mu}\frac{dx^\alpha}{dS}= \frac{1}{Me}g_{\alpha\mu}\frac{%
dx^\alpha}{d\lambda},  \label{23}
\end{equation}
where $\lambda$ is a parameter defined on the worldline of
particle on the brane. Now inserting this special $4$-vector into
equation (\ref{21}) and using equation (\ref{11b}) the result is
\begin{equation}
d\left(g_{\mu\nu}u^\mu u^\nu\right) = 2u^\mu u^4 K_{\mu\nu}dx^\nu + \frac{%
\dot{e}}{e}d\lambda = -2K_{\mu\nu}u^\mu u^4 dx^\nu,  \label{24}
\end{equation}
which is similar to the result obtained in (\ref{12}). Note that
for this special kind of the 4-vector, if the test particle does
not have an extra velocity component, like that of brane models
where test particles are confined to the brane, then according to
the above equation the length of the 4-velocity becomes
integrable,i.e. $dl=0$. But in general the induced vector
corresponding to the confined particle (in the case of RS brane
models) can have a component along the extra dimension. For
example, the pointlike gyroscopes are confined to the fixed brane,
but they  may have in general spin components along the extra
dimension and consequently, according to equation (\ref{21}) the
norm of spin is a non-integrable quantity. This is a another
reason that gyroscopes have a different behavior than spinless
test particles in brane models \cite{Seahra}. In the
non-integrable geometry, there is a well known method to measure
the ``length curvature'' ${\cal F} := d{\cal A}$ by means of the
so-called ``second clock effect''. Let us assume that, we have two
standard clocks which are close together and synchronized in the
beginning. Now if these two clocks are separated for a while and
brought together again later, they will be out of synchronization
in general. This is a well known effect from general and special
relativity and called ``first clock effect'' and often called the
twin paradox. The second clock effect exists if, in addition, the
units of the two clocks are deferent after their meeting again. In
Lorentzian spacetime there is no second clock effect for standard
clocks. Assuming that atomic clocks are standard clocks, then in
general, after the above argument, they have different properties.
To solve this problem Dirac \cite{Dirac} assumed that in practice
we have two different intervals: $ds_A$ and $ds_E$. The interval
$ds_A$ is referred to atomic units; it is not affected by ${\cal
A}$. The Einstein interval $ds_E$ is associated with the field
equations and the non-integrable geometry. Another solution to the
problem was given by Wood and Papini \cite{Papini}. In their
approach, the atom appears as a bubble. Outside one has the
non-integrable spacetime, and on the boundary surface and in the
interior of the atom we have ${\cal A}_\mu = 0$. The static
spherically entity is filled with ``Dirac matter" satisfying
equation of state like cosmological constant. Finally the third
method is discussed by Audretsch \cite{Aud} and Flint
\cite{Flint}. In this approach, the above solutions are classified
as non-quantum-mechanical ways and we can set second clock effect
as a quantum effect.

\section{Induced mass and Mach's principle}

To obtain the $4D$ observable mass $m$ that an observer measures, Lagrangian
(\ref{1}) gives the momentum conjugate to ${\cal Z}_A$ as
\begin{equation}
P_A=\frac{1}{e}{\cal G}_{AB}\dot{{\cal Z}}^B,  \label{25}
\end{equation}
so from the line element (\ref{2}) we have
\begin{equation}
{\cal G}^{AB}P_AP_B - M^2=0.
\end{equation}
For a $4D$ observer the motion is described by the 4-momenta $p^\mu$ such
that
\begin{equation}
g_{\mu\nu}p^\mu p^\nu = m^2,  \label{26}
\end{equation}
where $m$ is the induced mass. In order to obtain the induced mass of test
particles we project the 5-momenta $P_A$ into four dimensions. Assuming that
this projection is done by the vielbeins ${\cal Z}^A_\mu$ then
\begin{equation}
p_\mu = P_A {\cal Z}^A_{,\mu} = P_\mu = \frac{1}{e}u_\mu.  \label{27}
\end{equation}
On the other hand, comparing the relations $g_{\mu\nu}u^\mu u^\nu
= l^2$ and (\ref{26}) we find
\begin{equation}
m = \frac{l}{e}.  \label{28}
\end{equation}
It is easy to show, using equations (\ref{11b}), (\ref{12}) and (\ref%
{28}) that
\begin{equation}
\frac{dm}{m} = -{\cal A}_\alpha dx^\alpha.  \label{29}
\end{equation}
The author of \cite{Ponce1} obtained the same result by using
Hamilton-Jacobi formalism, instead of the geodesic one, and showed
that this expression showing  variation of mass is independent of
the coordinates and any parameterization used along the motion. An
interesting question at this stage is that what is the relation of
induced non-integrability from extra dimension to the physical
quantities that a $4D$ macroscopic observer measures. In this
section we will try to show that from the point of view of $4D$
observer the non-integrability and variation of mass are related
to the mach's principle. Before concentrating on Mach's principle,
it would be necessary to make some of the concepts to be used more
transparent and clear in what follows. Let us then start by making
a quick look at the IMT gravity. This would help us to grasp the
salient points of our discussion more easily. In this theory, the
motivation for assuming the existence of large extra dimensions
was to achieve the unification of matter and geometry, {\it i.e.},
to obtain the properties of matter as a consequence of extra
dimensions. In the IMT approach, Einstein equations in the bulk
are written in the form \cite{Wesson1}
\begin{equation}
{\cal R}_{AB}=0,  \label{30}
\end{equation}%
where ${\cal R}_{AB}$ is the Ricci tensor of the $5D$ bulk space. To obtain
the effective field equations in $4D$, let us start by contracting the
Gauss-Codazzi equations \cite{Eisen} \footnote{%
Eisenhart's convention \cite{Eisen} has been used in defining the Riemann
tensor.}
\begin{equation}
R_{\alpha \beta \gamma \delta }=2K_{\gamma [\alpha }K_{\beta ]\delta }+{\cal %
R}_{ABCD}{\cal Z}_{,\alpha }^{A}{\cal Z}_{,\beta }^{B}{\cal Z}_{,\gamma }^{C}%
{\cal Z}_{,\delta }^{D}
\end{equation}
and
\begin{equation}
2K_{\mu \lbrack \nu ;\rho ]}={\cal R}_{ABCD}{\cal Z}_{,\mu }^{A}{\cal N}^{B}%
{\cal Z}_{,\nu }^{C}{\cal Z}_{,\rho }^{D}.  \label{31}
\end{equation}%
where ${\cal R}_{ABCD}$ and $R_{\alpha \beta \gamma \delta }$ are
the Riemann curvature of the bulk and perturbed brane
respectively. To obtain the Ricci tensor and Ricci scaler of the
$4D$ brane we contract the Gauss equation. The result is
\begin{equation}
R_{\mu \nu }=(g^{\alpha \beta }K_{\mu \alpha}K_{\nu \beta}-KK_{\mu \nu})+%
{\cal R}_{ABCD}{\cal N}^{A}{\cal X}_{,\mu }^{B}{\cal X}_{,\nu }^{C}{\cal N}%
^{D},  \label{32}
\end{equation}
and
\begin{equation}
R={\cal R}+(K\circ K-KK)-2{\cal R}_{AB}{\cal N}^{A}{\cal N}^{B}+{\cal R}%
_{ABCD}{\cal N}^{A}{\cal N}^{B}{\cal N}^{C}{\cal N}^{D},  \label{33}
\end{equation}
where we have denoted $K\circ K := K_{\mu \nu }K^{\mu \nu }$ and $K :=
g^{\mu \nu }K_{\mu \nu }$. In the Gaussian form of the metric of the bulk
space, the last term appearing on the right hand side of equation (\ref{33})
vanishes. Using equations (\ref{32}) and (\ref{33}) we obtain the following
relation between the Einstein tensors of the bulk and brane \cite{Sasaki}
\begin{equation}
G_{AB}{\cal Z}_{,\mu }^{A}{\cal Z}_{,\nu }^{B}=G_{\mu \nu }-Q_{\mu \nu }-%
{\cal R}_{AB}{\cal N}^{A}{\cal N}^{B}g_{\mu \nu }+{\cal R}_{ABCD}{\cal N}^{A}%
{\cal Z}_{,\mu }^{B}{\cal Z}_{,\nu }^{C}{\cal N}^{D},  \label{34}
\end{equation}
where $G_{AB}$ and $G_{\mu \nu }$ are the Einstein tensors of the bulk and
the brane respectively, and
\begin{equation}
Q_{\mu \nu }=(K_{\mu }^{\,\,\,\,\,\,\gamma }K_{\gamma \nu }-KK_{\mu \nu})-%
\frac{1}{2}(K\circ K-K^2)g_{\mu \nu }.  \label{35}
\end{equation}
Now, decomposing the Riemann tensor of the bulk space into the Weyl and
Ricci tensors and Ricci scalar and using equation (\ref{30}), the Einstein
field equations induced on the brane become
\begin{equation}
G_{\mu \nu }=Q_{\mu \nu }-{\cal E}_{\mu \nu },  \label{36}
\end{equation}%
where ${\cal E}_{\mu \nu }={\cal C}_{ABCD}{\cal Z}_{,\mu }^{A}{\cal N}^{B}%
{\cal N}^{C}{\cal Z}_{,\nu }^{D}$ is the electric part of the Weyl Tensor of
the bulk space ${\cal C}_{ABCD}$. Note that directly from definition of $%
Q_{\mu \nu }$ it follows that it is independently a conserved
quantity, that is $Q_{\,\,\,\,\,\,;\mu }^{\mu \nu }=0$. All of the
above quantities in equation (\ref{36}) are obtained in the
perturbed brane since, according to the second equation of motion
(\ref{6}) the matter can not exactly be confined to the original
non perturbed brane. Hence from a $4D$ point of view, the empty
$5D$ equations look like Einstein equations with induced matter.
The electric part of the Weyl tensor is well known from the brane
point of view. It describes a traceless matter, denoted by dark
radiation or Weyl matter \cite{Padilla}. As was mentioned before,
$Q_{\mu \nu }$ is a conserved quantity which, according to the
spirit of the IMT theory should be related to the ordinary matter
as partly having a geometrical origin
\begin{equation}
Q_{\mu\nu} = - 8\pi G_N T_{\mu\nu}.  \label{37}
\end{equation}
This completes the description of the induced Einstein equation on
the perturbed brane. Now we are ready to discuss the physical
meaning of the variation of mass and non-integrability. In general
relativity we deal with large scales or at least up to the scales
in the order of millimeter. According to  \cite{Jalal2} the
influence of matter fields on the bulk space is small and at the
large scales the matter ``seems'' to be on the original brane
$\bar{g}_{\mu\nu}$. In what follows, we parameterize the path of a
particle with an affine parameter in the original brane. According
to definition (\ref {13}) we have
\begin{equation}
{\cal A}_\alpha = - K_{\alpha\beta}\bar{u}^4 \bar{u}^\beta (\frac{d\bar{s}}{%
ds})^2,  \label{38}
\end{equation}
where $ds^2=g_{\mu\nu}dx^\mu dx^\nu$ and
$d\bar{s}^2=\bar{g}_{\mu\nu}dx^\mu dx^\nu$ denotes the line
elements of the perturbed and original brane respectively and
$\bar{u}^\alpha$ is the $4$-velocity of the
test particle in the original non-perturbed brane. Now using equation (\ref%
{G}) up to order in $\xi$ we have
\begin{equation}
\left(\frac{ds}{d\bar{s}}\right)^2 = 1 + 2\xi
\bar{K}_{\mu\nu}\bar{u}^\mu \bar{u}^\nu + {\cal O}(\xi^2),
\label{39}
\end{equation}
and consequently inserting equation (\ref{39}) and (\ref{J}) into equation (%
\ref{38}) we obtain
\begin{equation}
{\cal A}_\alpha = -(\bar{K}_{\alpha\beta} - \xi \bar{K}_{\alpha \gamma}\bar{K%
}^\gamma_\beta)\bar{u}^\beta \bar{u}^4 + {\cal O}(\xi^2).  \label{40}
\end{equation}
The above equation together with equation (\ref{37}) leads us to
\begin{equation}
{\cal A}_\alpha dx^\alpha = -\bar{K}_{\alpha\beta}{\bar u}^\alpha {\bar u}%
^\beta \bar{u}^4 d\bar{s} - 8\pi G_N\xi \bar{u}^4(\bar{T}_{\mu\nu}\bar{u}%
^\mu \bar{u}^\nu + \frac{1}{2}\bar{T})d\bar{s},  \label{41}
\end{equation}
where $\bar{T}_{\mu\nu}$ is the energy-momentum tensor induced on
the original brane and the quantity $1/R :=
\bar{K}_{\alpha\beta}\bar{u}^\alpha \bar{u}^\beta$ is the normal
curvature \cite{Eisen}. In fact the normal curvature is nothing
more than the higher dimensional generalization of the familiar
centripetal acceleration \cite{Seahra}. Note that extrinsic
curvature according to equations (\ref{35}) and (\ref{37}) is
related to the energy-momentum tensor of the matter field. There
is a great difference in this point between IMT and brane models.
In brane models, confining the matter field to the brane dictates
that it is necessary to consider a confining potential or
vanishing normal
curvature. Another difference appears in exposition for $\bar{Q}_{\mu\nu} = \bar{T}_{\mu\nu}\bar{u}%
^\mu \bar{u}^\nu + \frac{1}{2}\bar{T}$ that appears in equation
(\ref{41}). In the brane models this term can be related to the
distribution of X-cold dark matter \cite{Jalal}, wheras according
to the our discussion in IMT this term is proportional to the
distribution of ordinary matter field, and also for the reason of
non-confining of matter field to any fixed brane we do not need
for any confining potential or vanishing normal curvature. All of
above shows the physical meaning of variation of mass in these
approaches are different.\newline

In 1880, Mach pointed out that the inertia depends on the
distribution of matter in the universe. This is called Mach's
principle \cite{Mach}. Since Mach's principle is not contained in
general relativity this leads to a discussion of attempts to
derive Machian theories. However, it is not explained satisfactory
even in more complete theories such as the Brans-Dicke
scalar-tensor theory \cite{Brans}. In the framework of Wesson's
gravity, the inertia of particle, according to equations (\ref{41}) and (%
\ref{29}) is related to the large scale distribution of matter in
the universe. This relation can be an explanation of mach's
principle, that inertial forces should be generated by the motion
of a body relative to the bulk of induced matter in the universe.
In this framework the generalization of Mach's
principle to the length of $4$-vectors can be done by inserting equation (%
\ref{41}) into equation (\ref{12})
\begin{equation}
\frac{dl}{l} = -\bar{K}_{\alpha\beta}{\bar u}^\alpha {\bar u}^\beta \bar{u}%
^4 d\bar{s} - 8\pi G_N\xi \bar{u}^4(\bar{T}_{\mu\nu}\bar{u}^\mu \bar{u}^\nu
+ \frac{1}{2}\bar{T})d\bar{s}.  \label{42}
\end{equation}
The above equation shows that the variation of length of the
$4$-velocity is generated by the motion of the body relative to
the matter distribution in universe too. For more
comprehensibility, as a simple example, consider a flat $5D$ bulk,
containing a $4D$ spacetime filled with a dust matter field: $\bar{T}%
_{\mu\nu} = \rho \bar{u}_\mu \bar{u}_\nu$ where
$\bar{g}_{\mu\nu}\bar{u}^\mu \bar{u}^\nu = -1$ and $\rho$ is the
matter density. Applying this kind of matter field to equation
(\ref{37}) gives
\begin{equation}
\bar{K}o\bar{K}- \bar{K}^2 = 8\pi G_N \rho.  \label{43}
\end{equation}
which is an algebraic equation on $\bar{K}_{\mu\nu}$ with solution
\begin{equation}
\bar{K}_{\mu\nu} = \sqrt{\frac{2}{3}\pi G_N \rho}\bar{g}_{\mu\nu},
\label{44}
\end{equation}
so that the extrinsic curvature is in direct proportion to the square root
of the density. Inserting the above relation in equations (\ref{29})and (\ref%
{41}) gives
\begin{equation}
\frac{dm}{m} = -\left( \sqrt{\frac{2}{3}\pi G_N \rho} - 4\pi G_N \xi \rho
\right)\bar{u}^4 d\bar{s}.  \label{45}
\end{equation}
This equation explicitly shows the variation of mass of the test particle in
the worldline of particle under the effect of distribution of induced matter
field on the our $4D$ universe.\newline

Note that the variation of mass is a special case of non-integrability,
induced on $4D$ submanifold. To an examining eye, let us start with the
equation (\ref{21}) for $4$-momentum
\begin{equation}
d(g_{\mu\nu}p^\mu p^\nu) = 2 p^\mu p^4 K_{\mu\nu}dx^\nu.  \label{46}
\end{equation}
Then according to the equations (\ref{26}) and (\ref{27}) we have
\begin{equation}
dm^2= 2 K_{\mu\nu}p^\mu p^4 dx^\nu = \frac{2}{e^2}K_{\mu\nu}u^\mu u^4 dx^\nu.
\label{47}
\end{equation}
The result by using the equation (\ref{28}) becomes
\begin{equation}
dm^2 = 2 \frac{m^2}{l^2}K_{\mu\nu}u^\mu u^4 dx^\nu,  \label{48}
\end{equation}
that is equivalent to the (\ref{29}). In the IMT approach the
extrinsic curvature is related to the energy-momentum tensor of
the induced matter on the brane. This dictates that according to
the equation (\ref{21}), the non-Riemannian structure on the brane
from  the point of view of $4D$ observer is related to the matter
contain of the universe.

\section{Conclusions}

In this paper, we have discussed Mach's principle and
non-integrability in Wesson's Induced Matter Theory. suppose one
carries out an infinitesimal displacement of a vector in the bulk
space. According to IMT the bulk space has Riemannian structure
and hence the change in the length of vector in the bulk space
obviously vanishes. On the other hand, one discovers that the
length of an induced vector on the brane is no longer constant
under parallel displacement. This means that the brane or our $4D$
spacetime is not a Riemannian space. The mentioned change is
induced by the bulk space and involves the extrinsic curvature and
velocity of the test particle along the extra dimension. The
non-integrability of the induced $4D$ geometry justifies revising
the definition and properties of the induced acceleration in the
geodesic equation (\ref{10a}) of the test particle, so that the
orthogonal part of acceleration vanishes $f_\mu u^\mu = 0$.
Another result of non-integrability is the variation of mass
which, according to equation (\ref{48}) is equivalent to the
Mach's principle. Hence one can extend the mach's principle in IMT
so that from the point of view of the $4D$ observer,
non-integrability is a consequence of the distribution of matter
in the universe.

\section{Acknowledgment}

I wish to thank H. R. Sepangi for his careful reading of the article and for
his constructive comments.

\end{document}